# Pre-CAT: A web-based, graphical user-interface toolbox for preclinical CEST-MRI data processing and analysis

Jonah Weigand-Whittier[1], Samuel Rubin[1], Cindy Ayala[1], Mark Velasquez[1], Nikita Vladimirov[2], Hadas Avraham[2], Or Perlman[2,3], M. Roselle Abraham[4], Moriel H. Vandsburger[1,*]

**1** Department of Bioengineering, University of California Berkeley, Berkeley, CA, USA

**2** School of Biomedical Engineering, Tel Aviv University, Tel Aviv, Israel

**3** Sagol School of Neuroscience, Tel Aviv University, Tel Aviv, Israel

**4** UCSF School of Medicine, San Francisco, CA, USA

* Corresponding author:

**Name** Moriel H. Vandsburger

**Department** Department of Bioengineering

**Institute** University of California Berkeley

**Address** 284 Hearst Memorial Mining Building

Berkeley, CA 94270

USA

**Email** moriel@berkeley.edu

**Manuscript word count**: 2797 (max 2800)

**Abstract word count**: 217 (max 250)

## Abstract

**Purpose**: As interest in CEST-MRI grows, particularly in the preclinical setting, the necessity for standardized and easy-to-use acquisition and data analysis pipelines has become apparent. While vendors have increasingly introduced support for CEST acquisitions on both clinical and preclinical hardware, image post-processing and analysis pipelines remain siloed based on privately developed code. We aim to develop an easy-to-use, open-source graphical user interface toolbox for pre-clinical CEST-MRI data analysis (Preclinical CEST-MRI Analysis Tool; Pre-CAT), supporting multiple acquisition types, organ systems, and CEST contrast mechanisms.

**Methods**: Pre-CAT was developed in Python and utilizes the Numpy, Scipy, and Matplotlib libraries for data analysis and plotting. Inbuilt data processing steps include image loading, reconstruction, post-processing, and segmentation. Pre-CAT also supports data analysis for QUESP, CEST-MRF, and field mapping experiments using consensus protocols and methods. Pre-CAT's web interface and GUI were developed using Streamlit, an open-source Python framework. Pre-CAT is hosted and accessible online and can be downloaded for local installation to complete the data analysis pipeline in roughly one minute using modern hardware.

**Results**: Pre-CAT analysis pipelines for Z-spectroscopy, CEST-MRF, and quantitative CEST (QUESP/QUEST) are demonstrated.

**Conclusion**: With the introduction of Pre-CAT, we aim to standardize the preclinical CEST-MRI data analysis pipeline, fostering collaboration across research sites and reducing methodological redundancy. Pre-CAT is open-source and relatively modular, encouraging the addition of new methods and protocols.

## 1 INTRODUCTION

Chemical exchange saturation transfer (CEST) is a novel magnetic resonance imaging (MRI) contrast mechanism first introduced in 2000 by Balaban et al. (1–3). Over the past quarter-century, CEST-MRI research activity has grown steadily (4), with nearly 2000 peer-reviewed articles published since 2020 (per Scopus search). Recently, both the research community and major vendors have accelerated the adoption of standardized CEST protocols. Key milestones include the release of Pulseq-CEST (5), an extension of the Pulseq library (6) for rapid prototyping and simulation, the 2022 consensus recommendations for amide proton transfer-weighted (APTw) imaging at 3T (7), and the introduction of the first FDA-approved sequence for 3D APTw imaging by Philips in 2018.

While the acquisition and analysis of CEST-MR data in clinical settings are rapidly standardizing, preclinical CEST imaging still constitutes the majority of ongoing research. However, without the pressure of clinical standardization, the analysis of such data remains non-uniform across institutions. Although numerous effective approaches for CEST contrast quantification (8–11) and quantitative CEST imaging (12–16) have recently been introduced, most preclinically focused CEST research groups still employ custom MATLAB or Python scripts for Z-spectral fitting and analysis. While the field benefits from several comprehensive CEST analysis packages (17), accessibility remains a hurdle; many of these tools are legacy projects, require proprietary licenses (e.g., MATLAB), or operate without GUIs.

The current paradigm poses several issues: first, those interested in performing de novo CEST experiments in small animals are required to develop their own custom tools for data analysis; second, due to the sensitivity of least-squares curve fitting (e.g., QUEST/QUESP, two-step Lorentzian fitting) to user defined initial conditions and upper/lower bounds, it is possible for different groups to derive nonuniform results from the same methods and data; and finally, collaboration and adoption of novel methods is hampered by the lack of standardized tools and workflows. This is particularly acute for pre-clinical cardiac CEST imaging where a recently introduced plug-and-play method enables easier and more robust data acquisition but lacks an accompanying analytical pipeline (18).

These issues, along with the recent introduction of open-source analysis software for similarly emerging methods such as XIPline (19) and the Hyperpolarized MRI Toolbox (20) for hyperpolarized $^{129}$Xe and $^{13}$C MRI, respectively, inspired the development of the Preclinical CEST Analysis Toolbox (Pre-CAT). Pre-CAT is an open-source graphical user-interface (GUI) webapp with straightforward workflows for importing, reconstructing, processing, and displaying preclinical CEST data. The straightforward GUI supports complete preclinical CEST imaging study workflows. Pre-CAT was designed to be highly modular and easily editable, allowing for the addition of novel CEST contrasts, Z-spectral fitting routines, field-mapping techniques, and quantitative methods. As the first comprehensive and completely open-source framework for preclinical CEST data analysis, Pre-CAT can serve as a platform for collaboration and transparency within the rapidly expanding CEST community.

# 2 METHODS

## 2.1 Overview and layout

Pre-CAT was developed in Python using Streamlit, an open-source framework for designing and sharing interactive data applications. As such, Pre-CAT can be deployed using several different methods. First, Pre-CAT can be run directly from the source code (available on GitHub) for local data analysis. Second, Pre-CAT can be deployed on shared computing resources for centralized data storage, processing, and analysis. And finally, Pre-CAT can be deployed online (e.g., on a research group website).

The Pre-CAT pipeline operates in a sequential manner, illustrated in Figure 1. First, experiment types are selected from a set list. This list currently includes CEST (i.e., conventional Z-spectroscopy), $B_0$ mapping using water saturation shift referencing (WASSR) (21), $B_1$ mapping using the double-angle method (22), quantification of exchange rate using varying saturation power (QUESP) (23,24), and CEST magnetic resonance fingerprinting (CEST-MRF) (12,13). Multiple experiment types can be selected, provided the uploaded data contains all relevant acquisitions. During data entry, the user also selects options regarding readout type (rectilinear or radial), pre-processing steps, and saved data filenames (Figure 2). For Z-spectroscopy, custom CEST contrasts can be selected based on need. Finally, each set of images undergoes reconstruction, pre-processing, segmentation, and post-processing based on the acquisition and experiment types.

## 2.2 Image acquisition

Cardiac and liver CEST-MRI data were acquired on a 7T Bruker PharmaScan 70/16 (Bruker, Ettlingen, Germany) system using the method and parameters described by Weigand-Whittier et al. (18), and peak saturation $B_1$ of 1.1μT and 5.0μT respectively. Separately, phantoms were created containing 50mM creatine in phosphate-buffered saline, titrated to various pH values between 6.92 and 7.18 using 1M sodium hydroxide and hydrochloric acid. QUESP (23) and CEST-MRF (12,13) datasets were acquired using these phantoms and a single-shot, spin echo EPI sequence (matrix size = 64×64 , FOV = $30\times30mm^2$ , slice thickness = 1mm) (12,25,26).

All animal experiments were performed in accordance with the Institutional Animal Care and Use Committee guidelines.

## 2.3 Image reconstruction and pre-processing

Image reconstruction can be performed either online (using Bruker's ParaVision software) or offline in Pre-CAT. For cardiac CEST acquisitions using the ungated, radial method described by Weigand-Whittier et al. (18), offline reconstruction using (the open-source) BART is required (27). An optional retrospective respiratory motion correction step is included for radial acquisitions (Supplemental Figure 1). This method bins acquisition segments based on detected respiratory motion by summing projections across each segment and calculating a moving average, then setting a threshold based on the standard deviation of the moving average and identifying segments in which the sum of the projections fall below this threshold. The same number of projections are binned across all frequency offsets so as not to affect quantified CEST contrasts. The pre-processing pipeline also includes Z-spectral denoising using principal component analysis, as described by Breitling, et al. (28). Notably, this denoising method is applied globally (as opposed to on a tissue-specific basis) and employs Malinowski's empirical indicator function to select the optimal number of principal components. For relevant acquisitions, thermal drift correction is also applied (29,30). Finally, users are given the option to rotate and flip images prior to processing and display. Generally, this is a matter of user preference; however, automatic myocardial segmentation will not function correctly unless the image orientation follows radiological convention for cardiac short-axis images (31,32).

**2.4 Segmentation**

***Manual segmentation***

Pre-CAT includes support for manual segmentation with polygonal regions of interest (ROIs). Users can define any number of ROIs using this method and assign labels to each ROI, which will be maintained throughout each processing step.

***Semi-automated cardiac segmentation***

For cardiac CEST-MR data Pre-CAT calculates left ventricular (LV) myocardial segments from a single mid-ventricle image using the following workflow:

1. The user manually defines epicardial and endocardial boundaries and draws a line between LV insertion points from anterior to inferior.
2. A single LV mask is defined by subtracting the endocardial area from the epicardial area.
3. A centroid is calculated from the LV mask. Points within the mask and insertion points are offset by the centroid so that the center of the LV is set to (0,0).
4. The angle of the anterior insertion point, the inferior insertion point, and every other point within the LV myocardium is calculated using the arctangent function.

5. The coordinate system is rotated so that the inferior insertion point is aligned at an angle of 0°. The area between the inferior and anterior insertion points is defined as the septum. The remaining area is defined as the free wall.
6. The septum is divided into two segments (anteroseptal and inferoseptal). The free wall is divided into four segments (anterior, anterolateral, inferolateral, and inferior).
7. A key/value pair (label/mask) is assigned to each segment and used for the remaining steps in the processing pipeline.

### 2.5 Post-processing

Image post-processing steps are experiment-specific and are executed automatically after pre-processing and segmentation. For all experiment types, acquisition-specific parameters (e.g., saturation $B_1$, frequency offsets, saturation pulse times) are automatically extracted from ParaVision method files.

#### 2.5.1 CEST Z-spectroscopy

Pre-CAT calculates CEST contrasts using a two-step Lorentzian difference fitting method as described by Zaiss et al. (33,34). The Z-spectrum is expressed as a sum of Lorentzian functions:

$$Z(\Delta\omega) = 1 - \sum_{i=1}^{N} L_i \tag{1}$$

where each Lorentzian $L_i$ describes some contribution to the Z-spectrum (e.g., rNOE, amide, amine). Each Lorentzian is described by:

$$L_i = A_i \frac{\frac{{\gamma_i}^2}{4}}{\frac{{\gamma_i}^2}{4} + (\Delta\omega - \omega_i)^2} \tag{2}$$

where $A_i$, $\gamma_i$, and $\omega_\mathrm{i}$ represent peak amplitude, full width at half-maximum (FWHM), and frequency offset relative to water, respectively. Lorentzian fits are calculated using a least-squares curve-fitting algorithm implemented in the Scipy library (35). First, water and magnetization transfer (MT) fits are calculated based on the entire Z-spectrum. Then, the water fit is used to correct for any $B_0$ inhomogeneity by subtracting the fitted water peak frequency offset from the uncorrected frequency axis. Water and MT fits are then calculated again using the corrected frequency axis, with spectral regions within the ±1.4-4 ppm bands excluded. Finally, these fits are subtracted from the total Z-spectrum to calculate the Lorentzian difference. Remaining CEST and rNOE contrasts are fitted to this Lorentzian difference. For each fitting step, fit root mean square error is also calculated and stored. This Lorentzian fitting process can be applied per manually defined ROI and/or on a pixelwise basis, depending on user selection.

#### 2.5.2 Field mapping

***$B_0$ mapping***

$B_0$ maps are calculated from WASSR acquisitions, as described by Kim, et al. (21). First, acquired WASSR spectra are interpolated over 1000 points using a cubic spline function. Then, a single Lorentzian is fitted to the WASSR spectrum by Equations 1 and 2. The frequency offset of the fitted peak is defined as the $B_0$ shift at each pixel. $B_0$ maps can be calculated either over user-defined ROIs or across the entire image.

***$B_1$ mapping***

$B_1$ maps are calculated using the double-angle method (22), where the true flip angle, $\theta$, at each voxel is given by:

$$\theta = \cos^{-1}\left(\frac{M_{2\theta}}{2M_{\theta}}\right) \tag{3}$$

where $M_\theta$ and $M_{2\theta}$ are image pixel values for acquisitions with flip angles $\theta$ and $2\theta$, respectively.
The relative flip angle scaling factor, $\kappa$, is then calculated from the real and nominal flip angle:

$$\kappa = \frac{\theta}{\theta_{nominal}} \tag{4}$$

If the matrix size of the acquired double-angle images is smaller than that of the reference image, Pre-CAT will match the size of the $B_1$ map to the size of the reference image using spline interpolation.

#### 2.5.3 Quantitative CEST

***QUESP***

Pre-CAT is capable of calculating quantitative solute proton volume fraction (concentration) and exchange rate maps using QUESP (23,24). The QUESP method takes advantage of the fact that the saturation efficiency, $\alpha$, is a function of saturation $B_1$:

$$\alpha = \frac{\omega_1^2}{\omega_1^2 + k_b^2} \tag{5}$$

where $k_b$ is the solute-water exchange rate and $\omega_1 = \gamma B_1$.
For CEST-weighted images acquired under continuous wave (CW) saturation, an analytical solution for $MTR_{asym}$ can be derived from the Bloch-McConnell equations:

$$MTR_{asym}\left(\alpha(B_1),\, t_p\right) = \frac{f_b k_b \cdot \alpha}{R_{1a} + f_b k_b \cdot \alpha} + (Z_i - 1)e^{-R_{1a}t_p} - \left(Z_i - \frac{R_{1a}}{R_{1a} + f_b k_b \cdot \alpha}\right)e^{-(R_{1a} + f_b k_b \cdot \alpha)t_p} \tag{6}$$

Where $f_b$ is solute proton volume fraction, $R_{1a}$ is the longitudinal relaxation rate for water, $Z_i$ is the initial longitudinal magnetization, and $t_p$ is the saturation pulse time.

Under steady-state saturation, when $t_p >> T_{1a}$, and when the longitudinal magnetization is allowed to fully relax between each readout, the inverse $MTR_{asym}$, $MTR_{Rex}$, can also be used to fit quantitative CEST parameters:

$$MTR_{Rex}\left(\alpha(B_1)\right) = \frac{1}{Z_{lab}} - \frac{1}{Z_{ref}} = \frac{1}{R_{1a}} f_b k_b \cdot \alpha \tag{7}$$

This method can be advantageous, as it naturally eliminates spillover and semisolid MT.

Finally, by rearranging the $MTR_{Rex}$ equation, the omega plot method can be used to plot the CEST intensity as a linear function of $1/\omega_1^2$:

$$y\left(\frac{1}{\omega_1^2}\right) = \frac{R_{1a}}{f_b k_b} + \frac{R_{1a} k_b}{f_b} \cdot \frac{1}{\omega_1^2} \tag{8}$$

Pre-CAT is capable of deriving quantitative CEST parameter maps using each of these methods, and will deliver a warning if, e.g., saturation steady-state conditions are not met by the extracted acquisition parameters.

To calculate quantitative parameter maps, Pre-CAT fits acquired data to one of the analytical solutions using a least-squares curve-fitting algorithm implemented in the Scipy library (35). $R_{1a}$ values are derived from either a single, user-defined $T_1$ value or from a pixelwise $T_1$ map acquired using the variable TR RARE method.

***CEST-MRF***

Pre-CAT is also capable of producing quantitative CEST parameter maps using CEST-MRF dot-product dictionary matching. This functionality relies upon the open-source open-py-cest-mrf library published by Vladimirov, et al. (13). The workflow for CEST-MRF dictionary generation and matching is displayed in Figure 3. Synthetic signals are generated from a CEST scenario configuration file, in which parameters such as $B_0$, a range of proton volume fractions and exchange rates for each solute pool, $R_{1b}$, $R_{2b}$, and a range of values for water relaxation rates ($R_{1a}$, $R_{2a}$) are defined. A script then automatically generates a Pulseq .seq object from acquisition parameters, and uses this sequence along with the configuration file to simulate the signal evolution for every combination of CEST and water relaxation parameters (5,6). During dictionary matching, the optimal set of CEST parameters is found by comparing the dot-products of synthetic signals within the dictionary against real, measured signal trajectories:

$$S(f_b, k_b, R_{1a}, R_{2a}) = \text{argmax}\left(\frac{\langle S_{synthetic}, S_{acquired}\rangle}{\|S_{synthetic}\| \cdot \|S_{acquired}\|}\right) \tag{9}$$

Signals are compared on a pixelwise basis within user-defined ROIs.

**3 RESULTS**

Results are returned to the user within the interface, with the option to download a zipped archive including all images, plots, and raw processed data. All plots and images are generated using the Seaborn (36) and Matplotlib (37) libraries.
Outputs vary based on experiment type, as shown in Figure 4. Processing times are shown in Table 1.

**3.1 Field mapping**

A representative results tab for field mapping (both $B_0$ and $B_1$) is shown in Figure 4b. Displayed results include a pixelwise $B_0$ map (either across the entire FOV or within an ROI, depending on user selection) and pixelwise $B_1$ maps (both across the entire FOV and interpolated on an anatomical reference). For field maps in the myocardium, boxplots displaying per-segment $B_0$ shifts and $B_1$ scaling factors are also displayed. The purpose of these per-segment results is to inform CEST contrast analysis (e.g., if minimal $B_0$ and $B_1$ inhomogeneities are observed in the anterior and anteroseptal segments, CEST contrast analyses should be constrained to these segments).

**3.2 Z-spectral and Lorentzian difference analyses**

A representative results tab for Z-spectroscopy in the murine heart is shown in Figure 4a. Displayed results include an ROI key, pixelwise CEST contrast maps (if selected), and per-segment fitted Z-spectra and Lorentzian difference plots.
Sample CEST contrast maps are shown for a genetically engineered mouse with a mutation leading to a phenotype that mimics human hypertrophic cardiomyopathy ($TnT^{+/-}$) and a corresponding littermate control (WT) (Supplemental Figure 2). Pre-CAT outputs facilitate additional pixel-wise filtering (e.g., by $\kappa$ and $B_0$ shift), enabling more physically accurate and representative CEST contrast analysis between groups by removing pixels with artificially attenuated CEST contrasts, such as partial volume effects resulting from motion dilution, as seen with the ungated method (Figure 5). This analysis approach can also be applied to additional organ systems, including the liver, as shown in Supplemental Figure 3.

**3.3 Quantitative CEST**

***QUESP***

A representative results tab for QUESP analysis is shown in Figure 4c. Displayed results include an ROI key, $T_1$ map, and calculated proton volume fraction, exchange rate, and fit $R^2$ maps. QUESP results also include a slider to adjust the range of percentile values displayed in parameter maps. This effectively prevents outliers (e.g., due to noise) from distorting the dynamic range of the parameter maps and can also be used to smooth parameter maps (e.g., in a phantom with clearly defined compartments).

***CEST-MRF***

A representative results tab for CEST-MRF is shown in Figure 4d. Displayed results include tabs for the calculated proton volume fraction, exchange rate, water relaxation rate, and dot-product maps. A table containing mean parameter values is also displayed. If the number of exchangeable protons is provided, Pre-CAT will automatically convert proton volume fractions to solute concentrations for display.

## 4 DISCUSSION

Pre-CAT is open-source via GitHub and designed to be highly modular. As a result, preclinical CEST-MRI researchers can easily contribute to the continued development of the application. By providing a basis for continued collaboration and sharing in the preclinical CEST space, Pre-CAT aims to reduce heterogeneity of analysis methods, accelerate multi-site studies, and provide an intuitive and easy-to-use CEST analysis platform for both experienced researchers and students.

### 4.1 Limitations and future work

The current release of Pre-CAT only supports single-slice CEST datasets acquired using sequences based on the MT module in ParaVision 6/7. We plan on adding support for ParaVision 360 sequences based on Bruker's new CEST module, multi-slice, and 3D acquisitions in a future release. We hope to see the functionality expanded to include additional custom contrasts, various novel contrast quantification techniques such as Polynomial and Lorentzian Fitting (PLOF) (8,9), multiple transmit field mapping methods such as Bloch-Siegert shift (38) and WASABI (39), and intelligent pixelwise mapping using techniques like image downsampling expedited adaptive least-squares (40). Several additions are currently forthcoming, including a GUI for Bloch-McConnell simulations and ratiometric analyses of creatine and phosphocreatine in the myocardium and skeletal muscle (41).

## 5 CONCLUSIONS

We have developed Pre-CAT as an open-source GUI-based tool for preclinical CEST-MRI data analysis. Using Pre-CAT, users can easily process and analyze entire CEST experiments both locally and online. We hope that Pre-CAT will become the preclinical standard for CEST data analysis, and that the preclinical CEST community will continue to collaborate towards making this tool as modular and robust as possible.

## ACKNOWLEDGEMENTS

This work was supported by the National Institutes of Health (NIH) grant *5UH3EB028908-04*, United States-Israel Binational Science Foundation (BSF), Jerusalem, Israel grant *2023139*, Department of Defense (DoD) grant *HT94252510620,* as well as the UC Berkeley France-Berkeley fund. The authors would like to acknowledge Jamie Enslein, who contributed to early code development and Dr. Moritz Zaiss for maintaining cest-sources.org.

**DATA AVAILABILITY STATEMENT**

Pre-CAT source code is available at https://github.com/jweigandwhittier/Pre-CAT. An online version of Pre-CAT is also hosted at https://proton.observer. The online version is currently password-protected and will be made available on an individual basis at the authors' discretion. Example datasets, including full cardiac (CEST Z-spectrum, WASSR, and double-angle B1 mapping) and variable pH creatine phantom (CEST-MRF) acquisitions, are available at 10.6084/m9.figshare.30505685.

## FIGURES AND TABLES

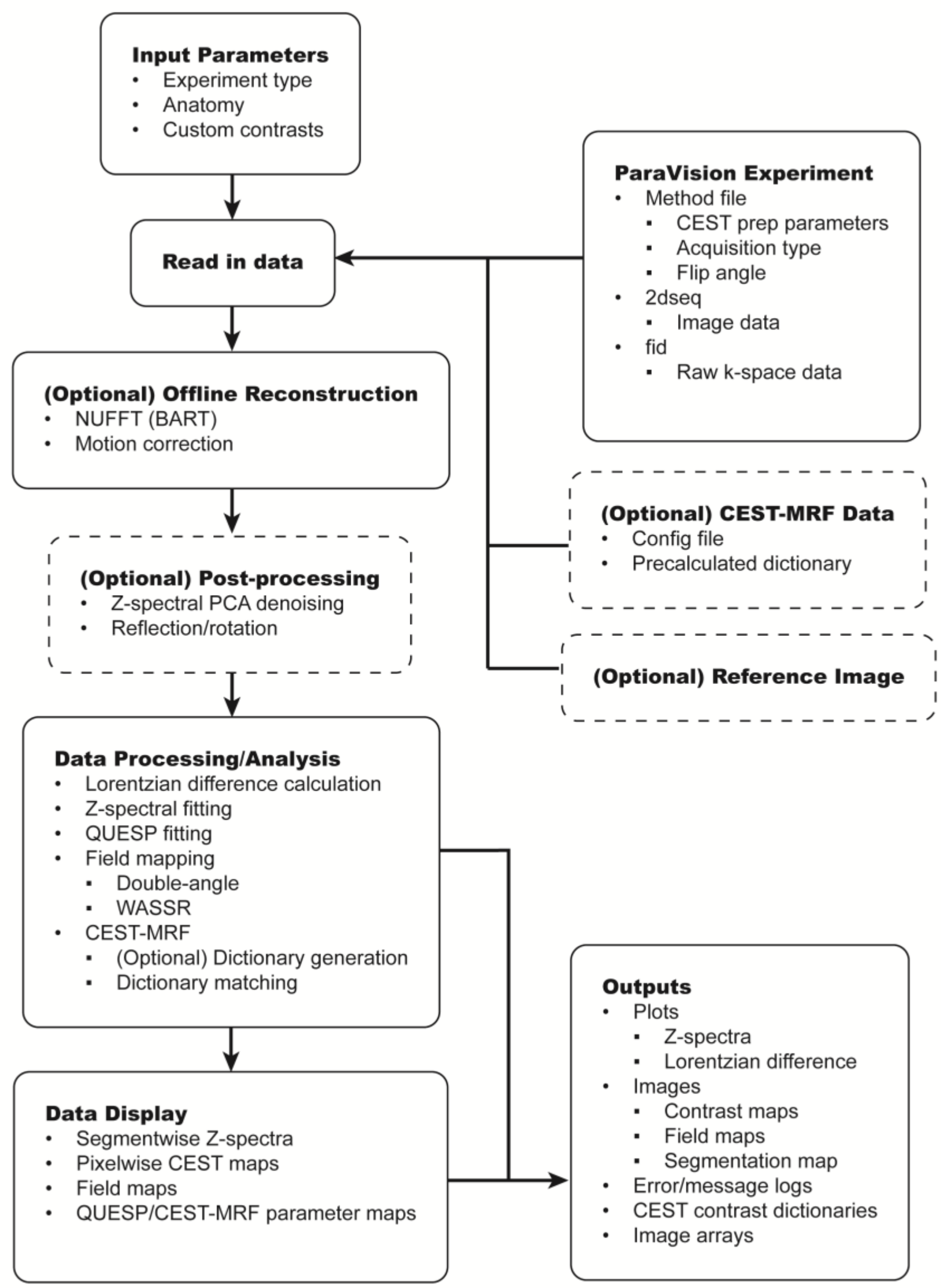

**Figure 1: Schematic diagram of the Pre-CAT data analysis pipeline.** This diagram illustrates the entire Pre-CAT pipeline including optional upload and data processing steps.

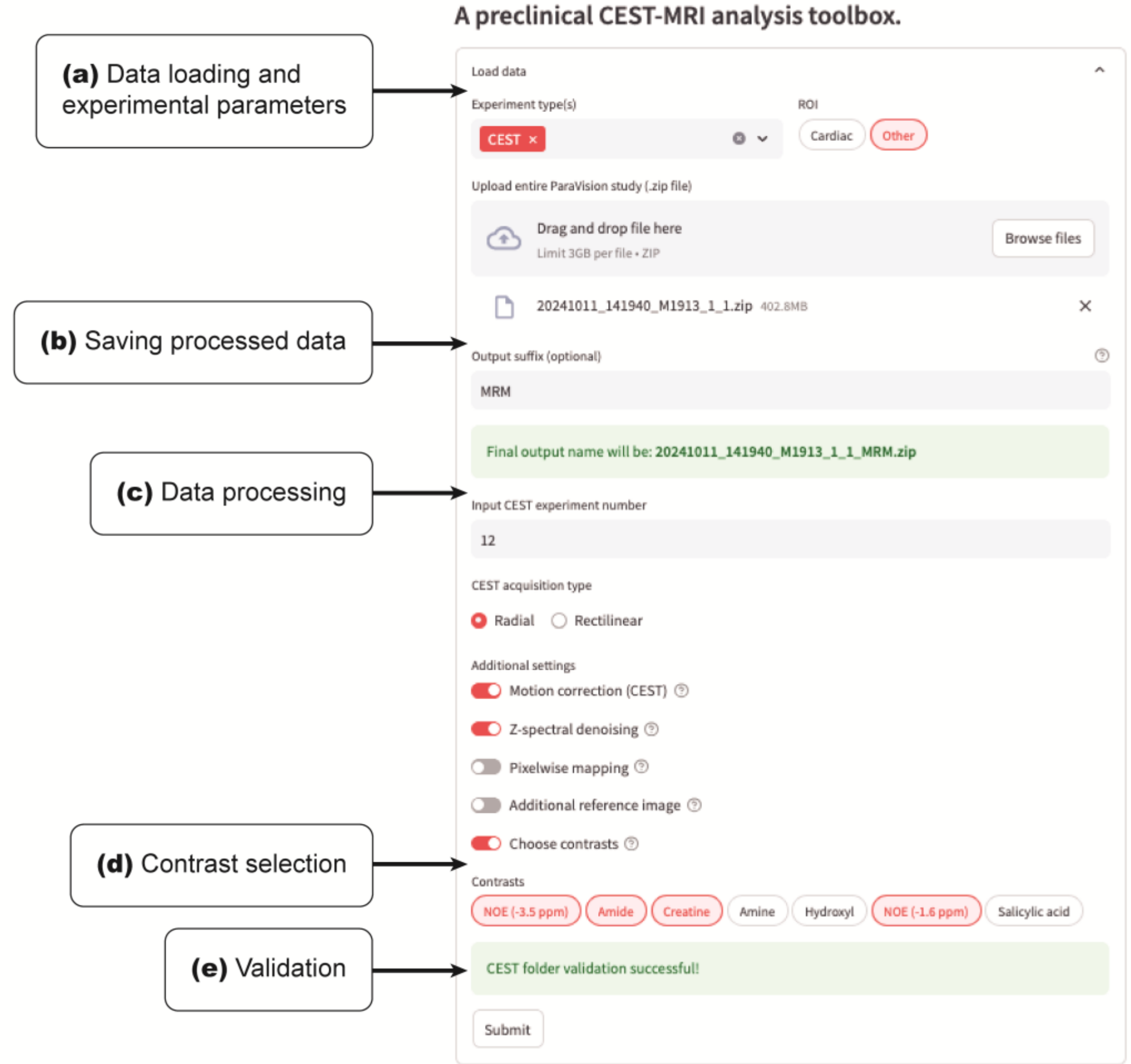

**Figure 2: Required and optional inputs and submission layout.** Users first select experiment type (e.g., CEST, WASSR, QUESP, CEST-MRF, etc.) and organ system/segmentation type (a). Users are then prompted to upload compressed ParaVision experiments and optionally add a suffix for downloadable processed data files (b). Each experiment is specified by the experiment number and acquisition/readout type. Additional settings, such as motion correction (for radial data), PCA Z-spectral denoising, pixel-wise mapping, and custom contrasts, are also selected (c). If custom contrasts are required, users can select from a predefined list (d) or define additional contrasts within the Pre-CAT code. Finally, all data is validated to ensure that it matches the selected experiment type and parameters (e).

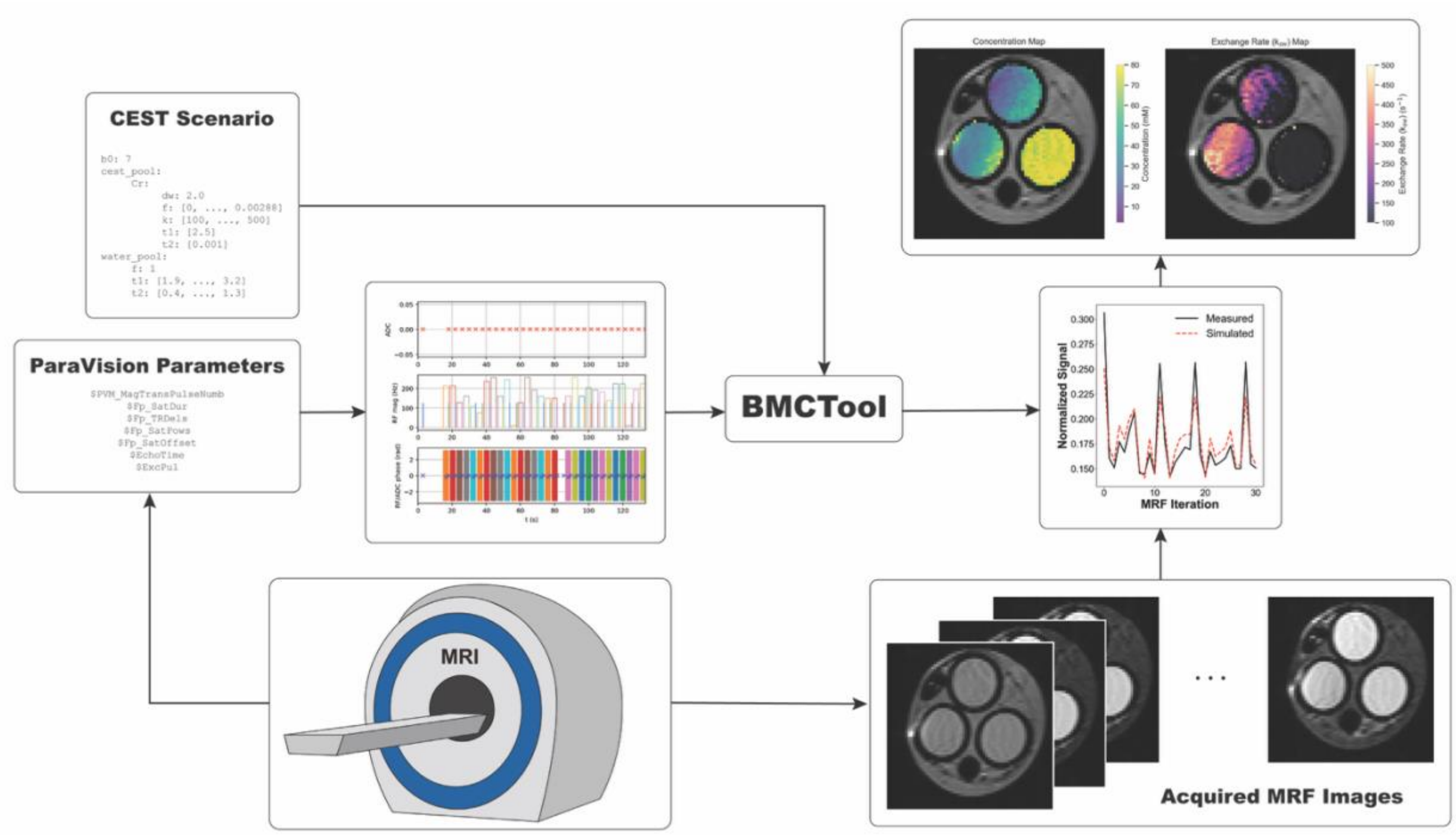

**Figure 3: Schematic diagram of CEST-MRF module.** First, a CEST scenario (i.e., field strength, range of water relaxation parameters, and solute parameter definitions) is defined by the user. Pre-CAT takes this scenario, along with a Pulseq sequence object automatically calculated from ParaVision scan parameters and uses a parallelized Bloch-McConnell simulator to generate a dictionary of signal trajectories for every combination of values defined in the CEST scenario. Finally, real signal trajectories from each pixel within the user-defined ROIs are matched to the simulated signal using the dot-product metric. The Pre-CAT CEST-MRF data analysis pipeline incorporates code from the recently published protocol by Vladimirov et al. (13)

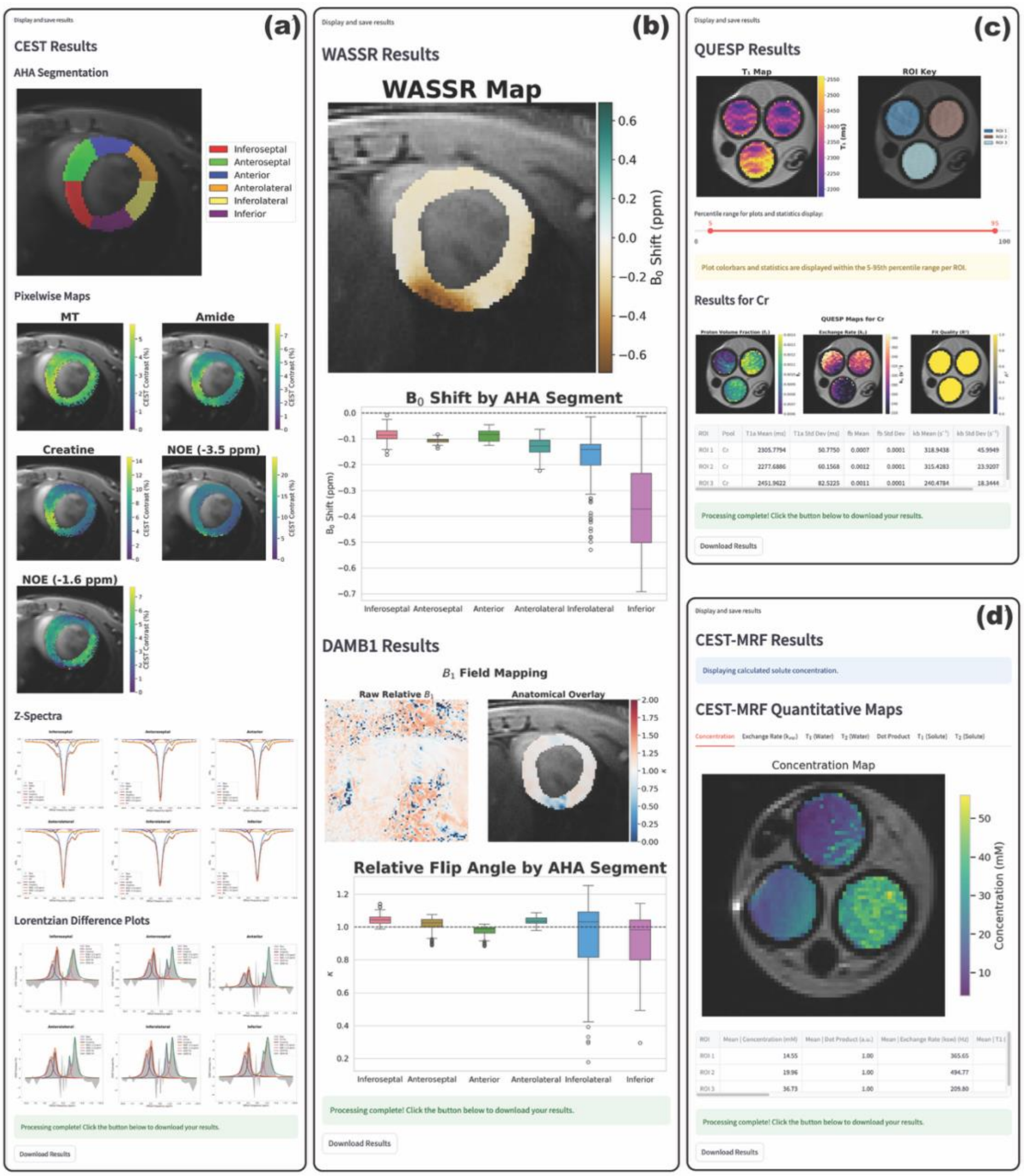

**Figure 4: Exemplary Pre-CAT output data.** Pre-CAT returns outputs based on the user-specified CEST experiment type. CEST Z-spectroscopy (a) outputs include pixelwise CEST contrast maps and segmentwise Z-spectra with Lorentzian difference plots. CEST-MRF (b) and QUESP (c) outputs include pixelwise quantitative proton volume fraction maps, exchange rate maps, and relevant water $T_1$/$T_2$ maps. Field map (d) outputs include pixelwise maps of $B_0$ shifts and $B_1$ scaling factor. For cardiac imaging, field map outputs also display field inhomogeneity by myocardial segment to aid in CEST contrast analysis. Segmentwise parameter tables are also output as CSV files. All experiment types output pickle files containing raw dictionaries and arrays generated during processing.

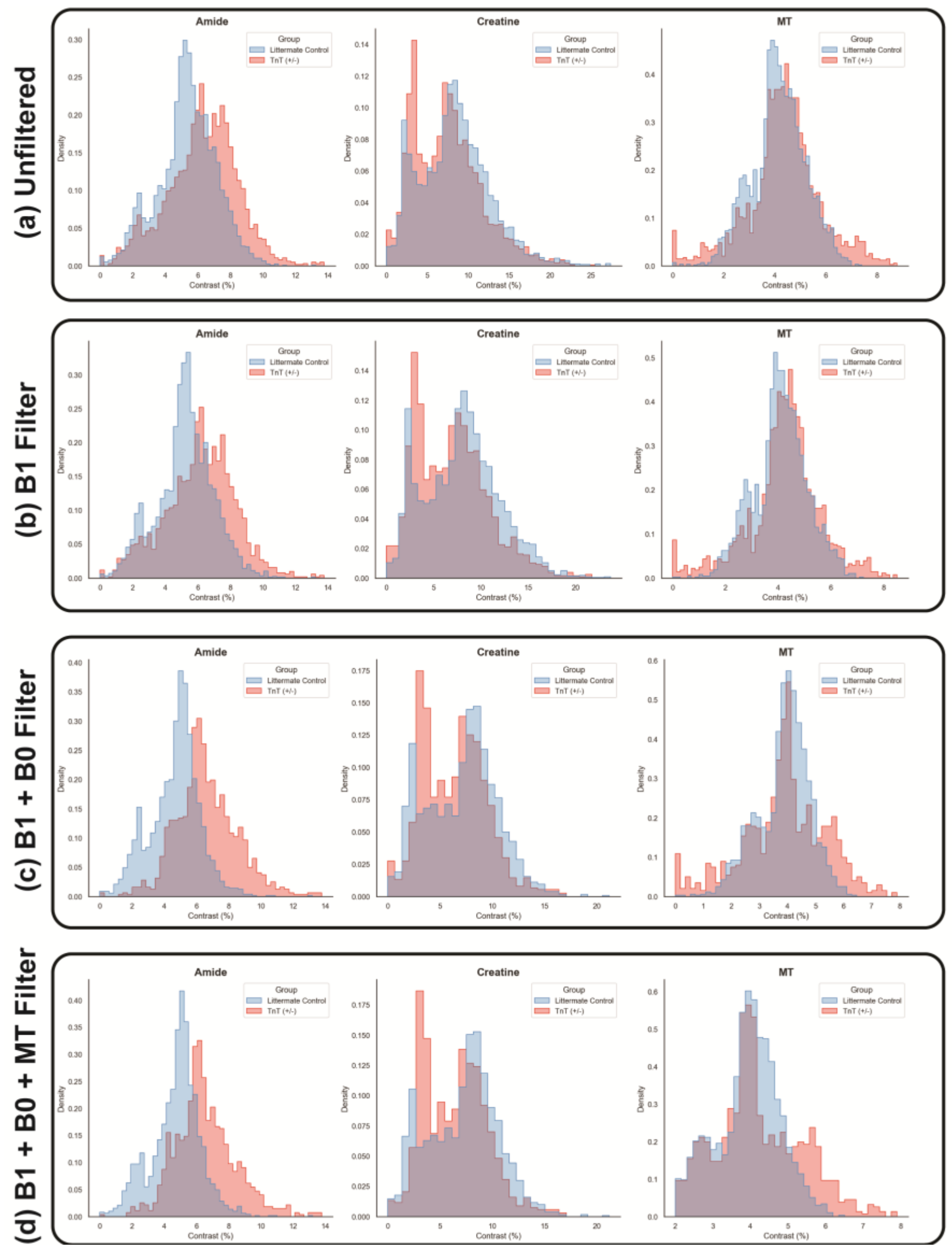

**Figure 5: Aggregated pixelwise CEST contrasts in a murine, transgenic model of hypertrophic cardiomyopathy ($TnT^{+/-}$) and a littermate control (WT) with progressive filtering steps.** Unfiltered, aggregated pixel data (a) exhibits subtle visual distinction in amide contrast distributions and bimodal creatine contrast distributions ($KS_{amide} = 2.37\times10^{-1}$ , $p = 4.09\times10^{-80}$). Iterative $\kappa$ (discard pixels with >1σ $\kappa$ per-animal) ($KS_{amide} = 2.56\times10^{-1}$ , $p = 3.00\times10^{-72}$) (b) and $B_0$ (discard pixels where $B_0$ shift > 0.25 ppm globally) ($KS_{amide} = 4.88\times10^{-1}$ , $p = 4.37\times10^{-156}$) (c) filtering narrows and separates amide contrast distributions, as reflected by the higher Kolmogorov-Smirnov (KS) test statistic.  A final MT filter (discard pixels with MT contrast < 2%) (d) effectively filters pixels that may exhibit lower CEST contrast due to, e.g., partial volume effects resulting from motion dilution, as seen with the ungated method ($KS_{amide} = 4.66\times10^{-1}$ , $p = 3.29\times10^{-129}$). This iterative filter process, facilitated by Pre-CAT output data, enables more physically accurate and representative CEST contrast analysis between groups.

| Processing Step | Computation Time (s) |
|---|---|
| Motion correction | 27.323 |
| PCA denoising | 0.023 |
| NUFFT reconstruction | 18.418 |
| Thermal drift correction | 0.478 |
| Segmentwise Lorentzian fitting | 0.061 |
| Pixelwise Lorentzian fitting | 27.091 |
| Pixelwise WASSR fitting | 3.656 |
| CEST-MRF sequence generation | 5.103 |
| CEST-MRF dictionary generation | 40.286 |
| CEST-MRF dot product matching | 3.523 |

**Table 1: Computation times for each step in the full reconstruction and data processing pipeline for exemplary CEST and CEST-MRF data.** Analysis was performed online using a Debian 13 server equipped with an Intel i5-12400 processor and 16GB of RAM. Computation times are reported for each step without user input. All CEST/WASSR fitting was performed over 858 pixels and 6 segments. CEST/WASSR image size was 192x192. CEST data contained 62 frequency offset images. CEST-MRF fitting was performed over 808 pixels. CEST-MRF data contained 31 images; dictionary length was 907,200 entries.

**SUPPLEMENTAL MATERIAL**

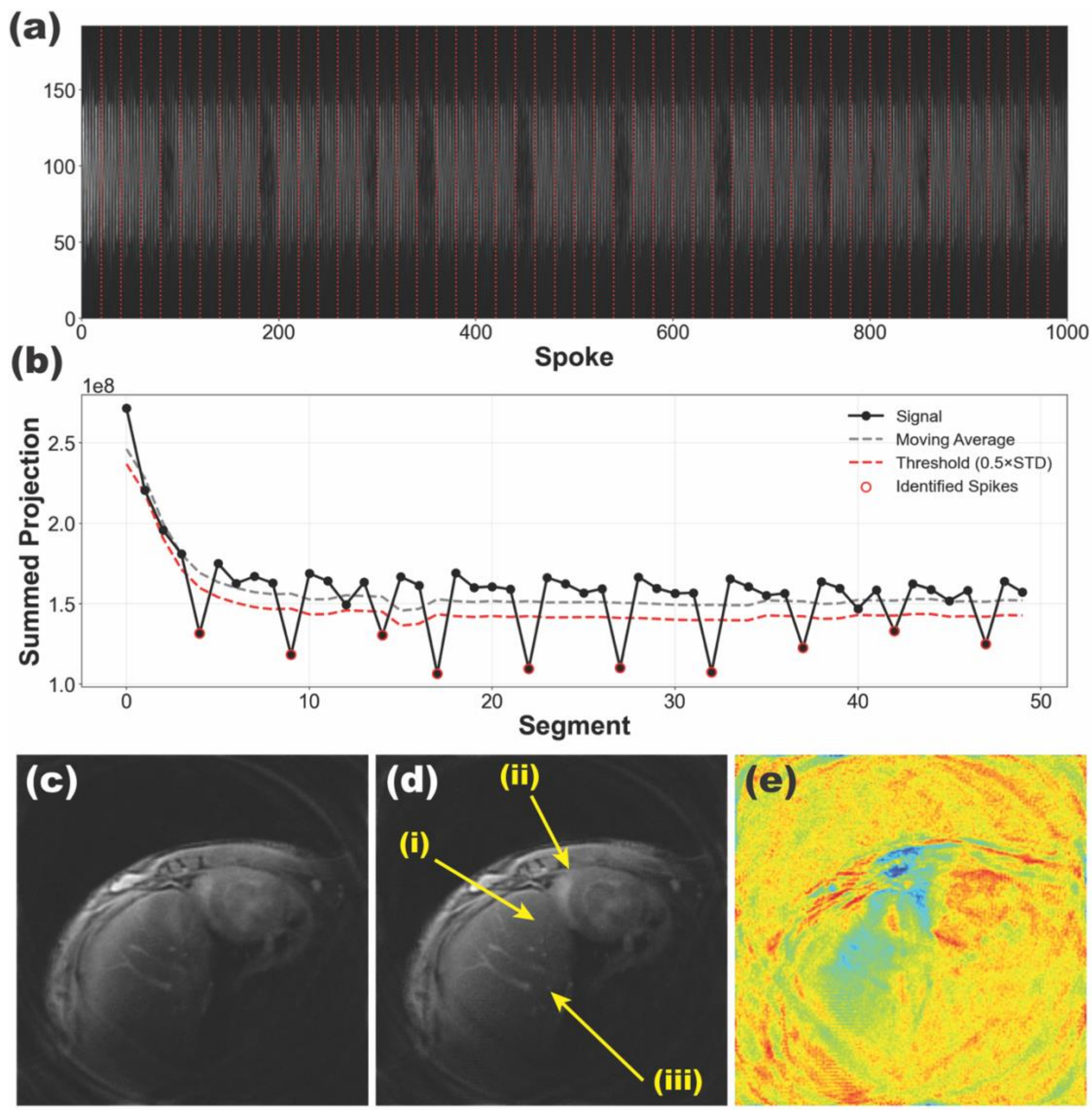

**Figure S1: A schematic illustrating the respiratory motion correction workflow.** Dark segments in the projection image (a) were identified as periods of respiratory motion. Projections were then summed over each segment, and a moving average was computed to quantitatively identify corrupted segments based on a signal drop threshold. The unfiltered image (c) exhibits blurring around the myocardium and liver. After filtering, the final image (d) demonstrated fewer susceptibility artifacts in the liver (i), improved edge sharpness in the LV and RV myocardium (ii), and enhanced visibility of hepatic vessels (iii). A difference map (raw - filtered) (e) was plotted using the Jet colormap to emphasize affected regions.

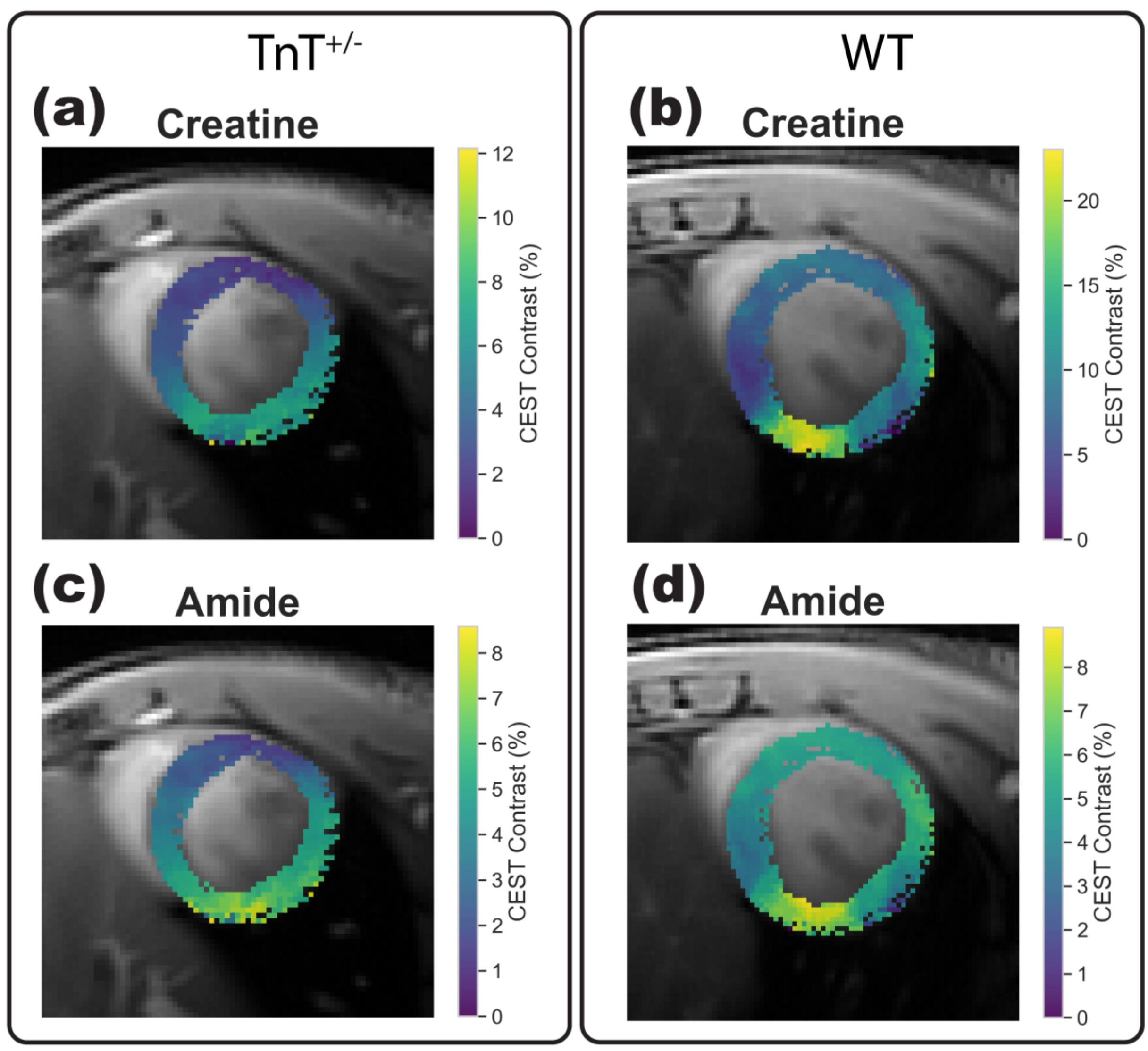

**Figure S2: Representative creatine and amide CEST contrast maps in a murine, transgenic model of hypertrophic cardiomyopathy (TnT$^{+/-}$) and a littermate control (WT).**

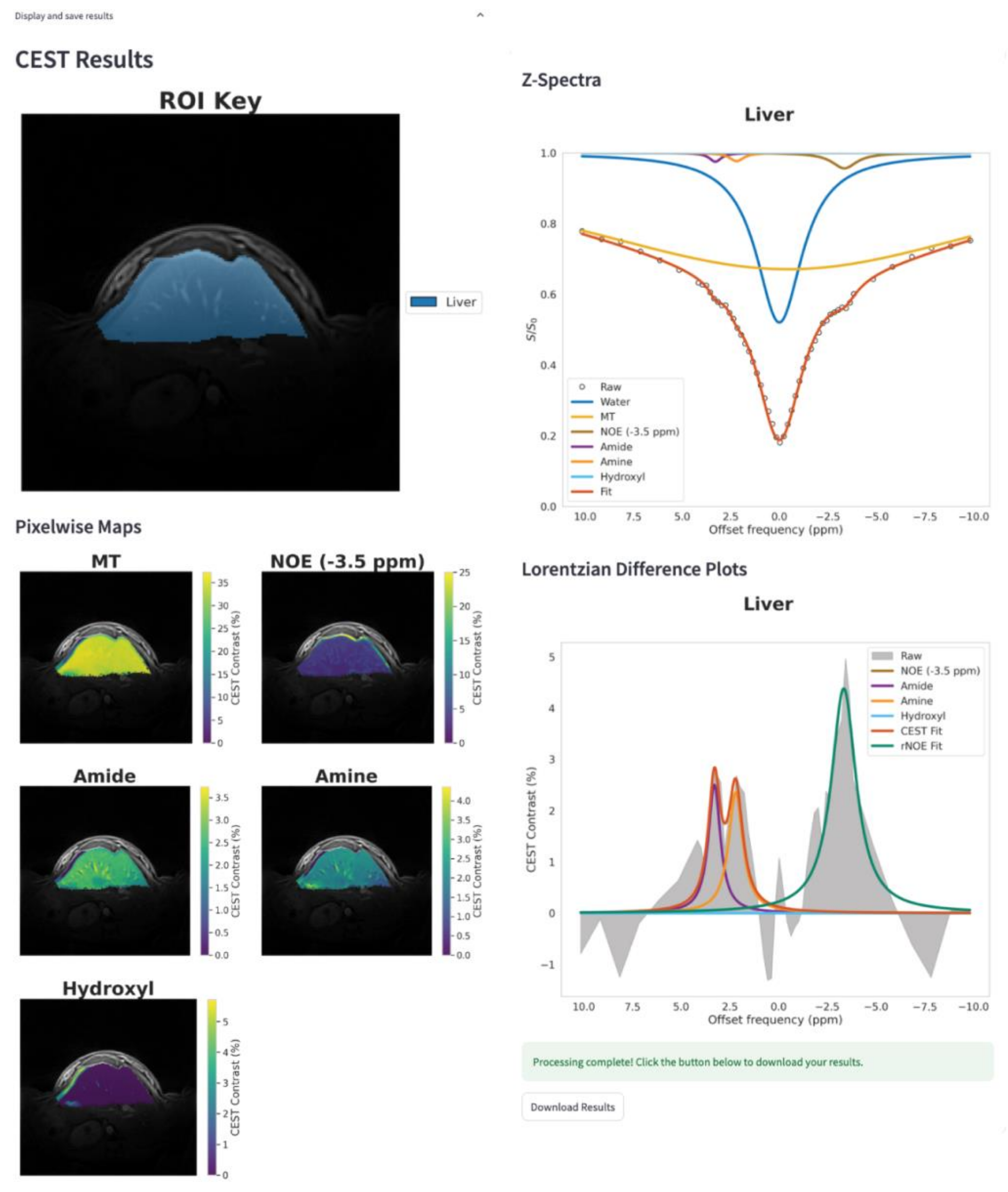

**Figure S3: Representative Pre-CAT output data for CEST imaging of the murine liver.** Outputs include pixelwise CEST contrast maps and segmentwise Z-spectra with Lorentzian difference plots. This animal was injected with a polyethylene glycol-based hydrogel, as seen in the hydroxyl contrast map.